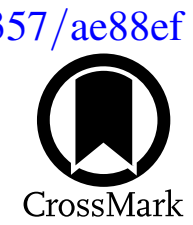

# An Approximate Bayesian Deep Learning Approach for Uncertainty-aware Differential Emission Measure Estimates in the Solar Corona from the SDO

N. Balodhi and R. J. Morton
School of Engineering, Physics and Mathematics, Northumbria University, NE1 8ST, UK

## Abstract

Accurately estimating the temperature distribution of solar coronal plasma, known as the Differential Emission Measure (DEM), is vital for understanding the thermodynamics of the corona and associated heating. However, recovering the DEM from multispectral observations like those from the Atmospheric Imaging Assembly (AIA) on board NASA's Solar Dynamics Observatory (SDO) is a mathematically ill-posed, underdetermined problem, and traditional regularization-based inversion methods are computationally intensive and provide limited uncertainty quantification. We present a deep learning framework for DEM reconstruction that incorporates Monte Carlo Dropout to perform approximate Bayesian inference, yielding per-pixel empirical distributions over the DEM that characterize epistemic or systematic model uncertainty in the learned inversion. The network is trained with a dual-head architecture supervising both AIA image reconstruction and DEM fidelity, with non-negativity enforced by construction. The network is trained directly on real SDO/AIA observations along with the associated DEM solutions from regularized inversion. We validate performance against both synthetic thermal distributions and real coronal data, demonstrating accurate recovery of thermal structure across a range of plasma conditions, while maintaining significant computational efficiency over traditional inversion techniques. This method ensures physically legitimate, non-negative solutions and provides per-pixel uncertainties, making it a reliable, high-speed, and uncertainty-aware tool for large-scale solar data analysis.

## 1. Introduction

The outermost layer of the Sun, the corona, is a region of intense, continuous activity hosting a variety of phenomena related to the interplay of the plasma and magnetic field (e.g., coronal jets, coronal mass ejections, and magnetohydrodynamic waves). Accurately estimating the temperature distribution of the plasma is essential in understanding the underlying evolution of plasma during such events, as well as the underlying heating mechanisms. Hence, obtaining reliable temperature estimates enables insights into the thermodynamics of the corona and the subsequent coronal heating problem (J. A. Klimchuk 2006; T. Wiegelmann et al. 2014; T. Van Doorsselaere et al. 2020).

The Solar Dynamics Observatory (SDO; W. D. Pesnell et al. 2012) is a NASA mission observing the Sun and solar activity that can have impacts on the Earth and the near-Earth space comprised of the upper atmosphere and the satellite orbiting region. The Atmospheric Imaging Assembly (AIA; J. R. Lemen et al. 2012) on board SDO is a four-telescope instrument that captures and provides high-resolution, full-disk images of the solar corona in 10 ultraviolet (UV), visible, and extreme ultraviolet (EUV) channels. The multispectral observations taken by AIA captures emission from ions formed at different temperatures in the corona and provides a wealth of information about the structure and evolution of the observed system. Further, with a 12 s cadence, AIA has collected millions of high-resolution images of the Sun since its launch in 2010. The temperature distribution of the plasma, however, cannot be directly obtained from the AIA images (P. Boerner et al. 2012). This is because the narrowband wavelength bandpass for each channel captures emissions from multiple ion species (B. O'Dwyer et al. 2010). Hence, the instrument response function for each channel is multithermal, and the contributions to emission within each bandpass span a range of temperatures and cannot be directly separated.

Figure 1 shows the wavelength-dependent response functions as a function of temperature for the EUV channels. The observed intensity in each channel can be inverted to provide estimates for the plasma emission at different temperatures. The Differential Emission Measure (DEM; $cm^{-5}\,K^{-1}$) is defined as the amount of plasma along a line of sight (LOS) that contributes to the radiation emitted by the solar corona within a temperature range $T + dT$ (I. J. D. Craig & J. C. Brown 1976; M. J. Aschwanden 2004; L. Golub et al. 2004):

$$\xi(T) = n_e n_{\mathrm{H}} \frac{dl}{dT}, \quad (1)$$

where $\xi(T)$ denotes the DEM value at temperature $T$. The quantities $n_e$ and $n_{\mathrm{H}}$ denote the electron and hydrogen number densities, respectively, and $l$ denotes the path length along the LOS. The corona is an optically thin plasma, and the column depths are typically hundreds of Mm (P. R. Young et al. 1999).

The intensities observed by the telescope in different wavelength channels can be determined by integrating the product of instrument response functions with the DEM distribution:

$$y_i = \int_0^\infty K_i(T)\xi(T)\,dT, \quad (2)$$

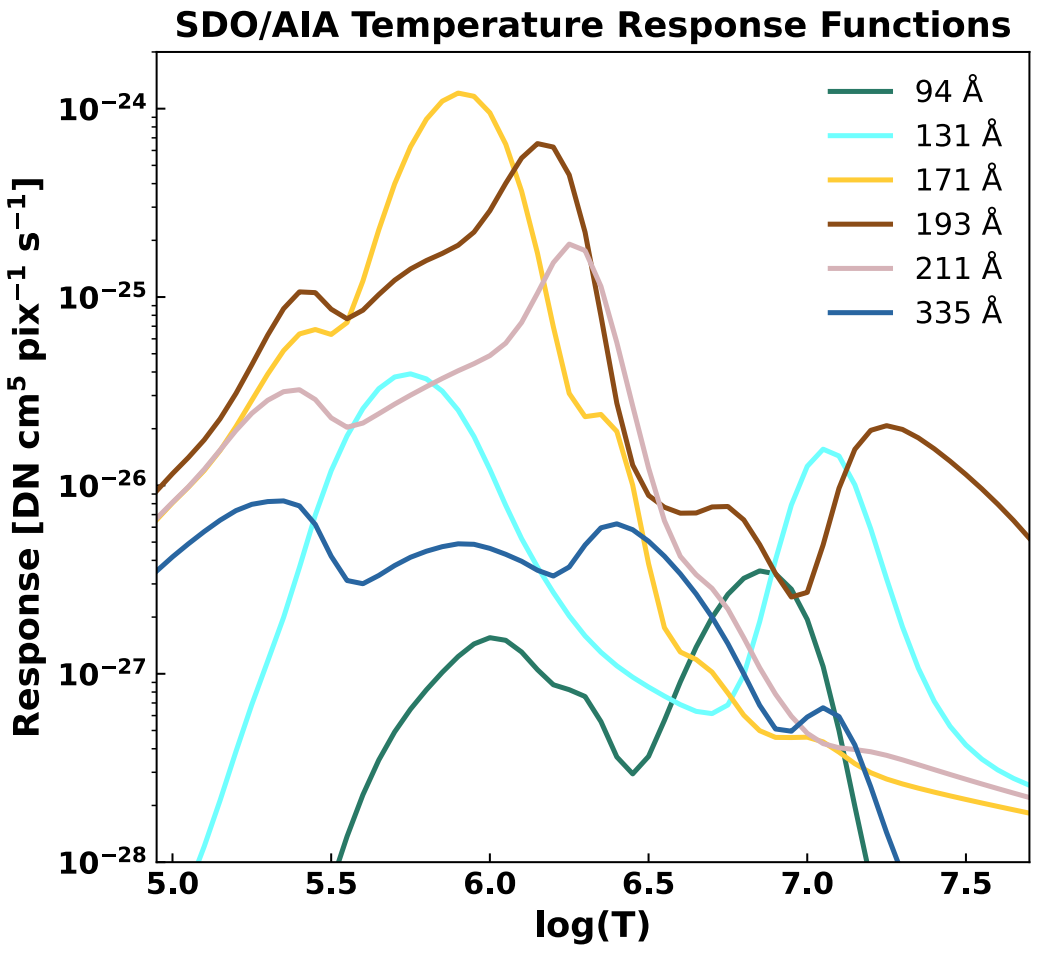

**Figure 1.** The response function of each wavelength channel for the AIA/SDO. The 304 Å is excluded from the plot and the analysis, as it is dominated by optically thick He II emission from the chromosphere. These response functions are computed using the *aia_get_response* (https://hesperia.gsfc.nasa.gov/ssw/sdo/aia/idl/response/aia_get_response.pro) function in SolarSoft (S. L. Freeland & B. N. Handy 1998).

where $y_i$ is the observed intensity of the image in the $i$th wavelength channel, and $K_i(T)$ is the temperature-dependent response function of the instrument for the $i$th channel. Note that the mapping of the temperature distribution to intensities, via Equation (2), is valid if the radiation is both optically thin and in thermal equilibrium. This is an example of an underdetermined system of equations that possesses more variables than unknowns and subsequently infinite solutions. The matrix representation of this can be given by

$$y_i = K_{i,j}\xi(T_j). \tag{3}$$

Mapping the plasma emission from the corona across a range of temperatures using the solar images can provide an estimate for the thermal structure and temperature distribution throughout the corona. Any straightforward attempt to reconstruct the temperature-dependent plasma emission from the intensity, however, leads to amplifications of errors and physically illegitimate solutions (in the form of negative or undefined emission values). Thus, reconstructing the DEM values requires the addition of external smoothing constraints to prevent the amplification of these errors and obtain the most probable set of solutions. Some common approaches to solving the DEM inversion problem involve the use of regularization, in particular the popular method of using the $L_1$ and the $L_2$ norms to provide regularization (I. G. Hannah & E. P. Kontar 2012; J. Plowman et al. 2013; M. C. M. Cheung et al. 2015). The use of these two norms can lead to differences between the inversion results, with the $L_1$ norm promoting sparsity within the solutions. Bayesian techniques have also been used to study coronal temperature distributions: e.g., H. Warren et al. (2016) uses a sparse Bayesian analysis to capture DEMs, and K. P. Dere (2022) utilizes Markov Chain Monte Carlo simulations to study DEMs and temperature distributions in the quiet and active solar corona (see also V. Kashyap & J. J. Drake 1998).

However, these methods often come with high time requirements and computational complexity, as well as other issues, including undefined solutions. Furthermore, it is important to account for uncertainties in the estimated DEMs, to be able to quantify the reliability of subsequent estimates of plasma parameters. P. J. Wright et al. (2019) proposes a deep learning approach to DEM inversion (called DeepEM) that takes full-disk images of the solar corona in optically thin EUV channels from the SDO/AIA instrument as inputs and generates DEM maps across a range of temperatures. DeepEM is trained using DEMs produced via the basis pursuit method. The neural network–based approach significantly reduces the time required to capture plasma emissions from observations and provides nonzero, positive DEM solutions. However, the network architecture used by P. J. Wright et al. (2019) can be viewed as a frequentist approach, estimating the maximum likelihood values for the network weights. In general, typical formulations of neural networks for regression problems are unable to provide uncertainties on the network parameters, and as a consequence, they cannot provide an estimate of uncertainty on the predictions.

Bayesian methods provide a natural way to incorporate uncertainty into neural networks. In general, the Bayesian approach to optimization provides a posterior probability distribution for individual parameters and the model as a whole. Once the posterior distribution is defined, it is possible to use it to obtain the push-forward[1] posterior distribution for deterministically derived quantities. This finds use because it is often of interest to estimate a representative temperature and density for the plasma from the DEM. Hence, once the posterior for the DEM has been obtained, it is straightforward to obtain posterior distributions for plasma parameters of interest. In practice, samples from the posterior are obtained to derive an empirical posterior distribution, via a Monte Carlo (MC) method, and these samples are used in calculating an empirical push-forward posterior.

In this paper, we expand upon the work of P. J. Wright et al. (2019) and present a regularized deep learning approach to solve the ill-posed DEM inversion problem. We utilize a simple approach referred to as the MC dropout, which effectively introduces Bayesian features into the network. We focus on applying this approach to estimating temperatures from optically thin AIA passbands. We test the ability of MC dropout to compute the uncertainties on the predicted DEM through the generation of synthetic observations where the underlying DEM distribution is known.

## 2. An Approximate Bayesian Method

Convolutional neural networks (CNNs; K. O'Shea & R. Nash 2015) are powerful tools for processing data resembling a grid structure, like images. Unlike artificial neural networks, they are spatially aware and use filters to identify nonlocal or neighboring pixel information to create feature maps from inputs and generate an output. However, CNNs are prone to overfitting by capturing noise and irrelevant details as part of the training data. This can be caused by having insufficient training data to capture the relationship accurately, or by using a network with too many free parameters, such that the network can mistake noise as underlying pattern information between the input and output. Dropout is a popular regularization technique used in neural networks (G. E. Hinton et al. 2012) that randomly removes certain hidden units from a network while training. This helps in creating a sparse network while training and reduces complexities within the model parameters, which effectively tackles overfitting to the training data and helps it generalize over previously unseen images. However, Y. Gal & Z. Ghahramani (2016) and Y. Gal (2016) demonstrate that including dropout and a weight decay term during training also admits a Bayesian interpretation (a more detailed discussion is given in the Appendix). Weight decay is another regularization technique that works by adding a penalty term to the loss function, $\mathcal{L}$, restricting large weights and shrinking them to near zero. The new loss function is given by $\mathcal{L}_{\rm reg} = \mathcal{L} + \lambda \sum w_i^2$, where $\lambda$ is the regularization term penalizing the network weights $w_i$.

[1] A push-forward distribution is the probability distribution of a random variable, $Y$, formed by applying a deterministic transformation, $g$, to another random variable, $X$, i.e., $Y = g(X)$.

In a Bayesian framework, rather than producing a single point estimate, one seeks the full predictive distribution for the output $\boldsymbol{y}^*$ (in our case, $\xi(T)$ given in Equation (2)), given test input $\boldsymbol{x}^*$ (the AIA channel observable given by $y_i$ in Equation (2)):

$$p(\boldsymbol{y}^*|\boldsymbol{x}^*, D) = \int p(\boldsymbol{y}^*|\boldsymbol{x}^*, \boldsymbol{w})\, p(\boldsymbol{w}|D)\, d\boldsymbol{w}, \tag{4}$$

where $p(\boldsymbol{y}^*|\boldsymbol{x}^*, \boldsymbol{w})$ is the likelihood, $\boldsymbol{w}$ are the model parameters (e.g., network weights), $D$ represents the training data, and $p(\boldsymbol{w}|D)$ is the posterior over parameters. Here, integration marginalizes over all possible parameter configurations weighted by their posterior probability, given training data $D$. This integral is analytically intractable for neural networks.

Y. Gal (2016) shows that a network trained with dropout applied after each layer, along with weight decay, is mathematically equivalent to performing approximate variational inference in a Bayesian neural network. Hence, the network is a probabilistic model in which weights are treated as random variables with a prior distribution, rather than fixed point estimates. The dropout operation provides the variational distribution, approximating the true posterior over network weights. From a practical perspective, running the trained network multiple times with dropout active, each time randomly deactivating a different subset of units, produces a distribution of predictions whose spread directly reflects the model's epistemic uncertainty. This procedure, MC dropout, therefore provides a computationally inexpensive approximation to Bayesian inference without modifying the training procedure.

MC dropout provides a practical approximation, where the network is evaluated $T$ times on the same input, each time with a different randomly sampled dropout mask, yielding the MC estimate

$$p(\boldsymbol{y}^*|\boldsymbol{x}^*, D) \approx \frac{1}{T}\sum_{t=1}^{T} p(\boldsymbol{y}^*|\boldsymbol{x}^*, \boldsymbol{w}_t). \tag{5}$$

The mean and variance of this ensemble directly provide the posterior predictive mean $\hat{\mu}_y$ and an estimate of the total predictive uncertainty, decomposed into a model (epistemic) component, reflecting uncertainty in the network weights, and a data noise (aleatoric) component intrinsic to the observations. In this work, we focus on the epistemic component.

## 3. Bayesian Network

### 3.1. Training Data

Constructing a neural network for our problem requires curating a training data set that allows learned mapping from the observed intensity in coronal channels to temperature-dependent emission measures. The training and testing data consist of SDO/AIA observations in optically thin EUV channels at 2010 September 17 UTC with low solar activity and at 2014 April 20 UTC and 2014 May 18 UTC, which marked the peak of the solar cycle and increased activity. These dates were chosen to ensure the training data set captures a range of differential emission profiles. Previous work on measuring the DEM of the coronal plasma demonstrates that emission from the quiet Sun has narrow DEM distributions, while those from active regions are broader (H. Morgan & Y. Taroyan 2017). The DEMs for these two regions show only minor variation over the solar cycle. Hence, using these three dates gives a good coverage of expected DEM profiles in the corona for non-flaring plasma.

We calculate the DEMs in 18 temperatures spanning a range of temperatures ($\log_{10} T = 5.5$–$7.2$) with a bin width of 0.1, using the regularized inversion ($L_2$) method (I. G. Hannah & E. P. Kontar 2012).[2] The training data are generated by dividing full-disk images of size 4098 × 4098 pixels$^2$ into 64 cutouts of size 512 × 512 pixels$^2$ each, to closely observe smaller regions of the corona containing quiet Sun, active regions, sunspots, or a combination of different coronal phenomena. The final data set has 192 cutouts of AIA observations (512 × 512 × 6) and their DEM maps (512 × 512 × 18), which are then divided into training, validation, and testing sets of 72, 32, and 24 images each. While the numbers of images are limited, each individual pixel acts as a unit of training data for the per-pixel mapping, resulting in a training sample size of over one million. All images used for training, validation, and testing have been preprocessed to Level 1.5, using version 0.7 of the *aiapy* open source software package (W. T. Barnes et al. 2020) in Python.

### 3.2. Model Architecture

To map EUV intensity to DEM values, we train a neural network that leverages the instrument response functions of the AIA. The network incorporates regularization through a combination of $L_2$ norm penalties with a regularization parameter of $10^{-5}$ and dropout layers. Dropout is actively enabled during inference to quantify model uncertainty, following a Bayesian approximation approach. It is important to note that these are regularization constraints within the network parameters that penalize our loss function, to help the model generalize better, while the training data set derived from I. G. Hannah & E. P. Kontar (2012) itself causes $L_2$ regularized coefficients within the DEMs prior to training.

The model architecture is designed to learn the mapping between EUV intensities and the underlying temperature distribution of the plasma as described by the DEM, effectively learning the inversion process constrained by $L_2$ priors. It then further uses the response function to simulate synthetic EUV observations from its predicted DEM outputs.

[2] We used the Python version based off the IDL code given in https://github.com/ianan/demreg. This method offers several hyperparameters in order to adjust the optimization. We do not provide an initial guess solution, and the hyperparameters are set to their default values.

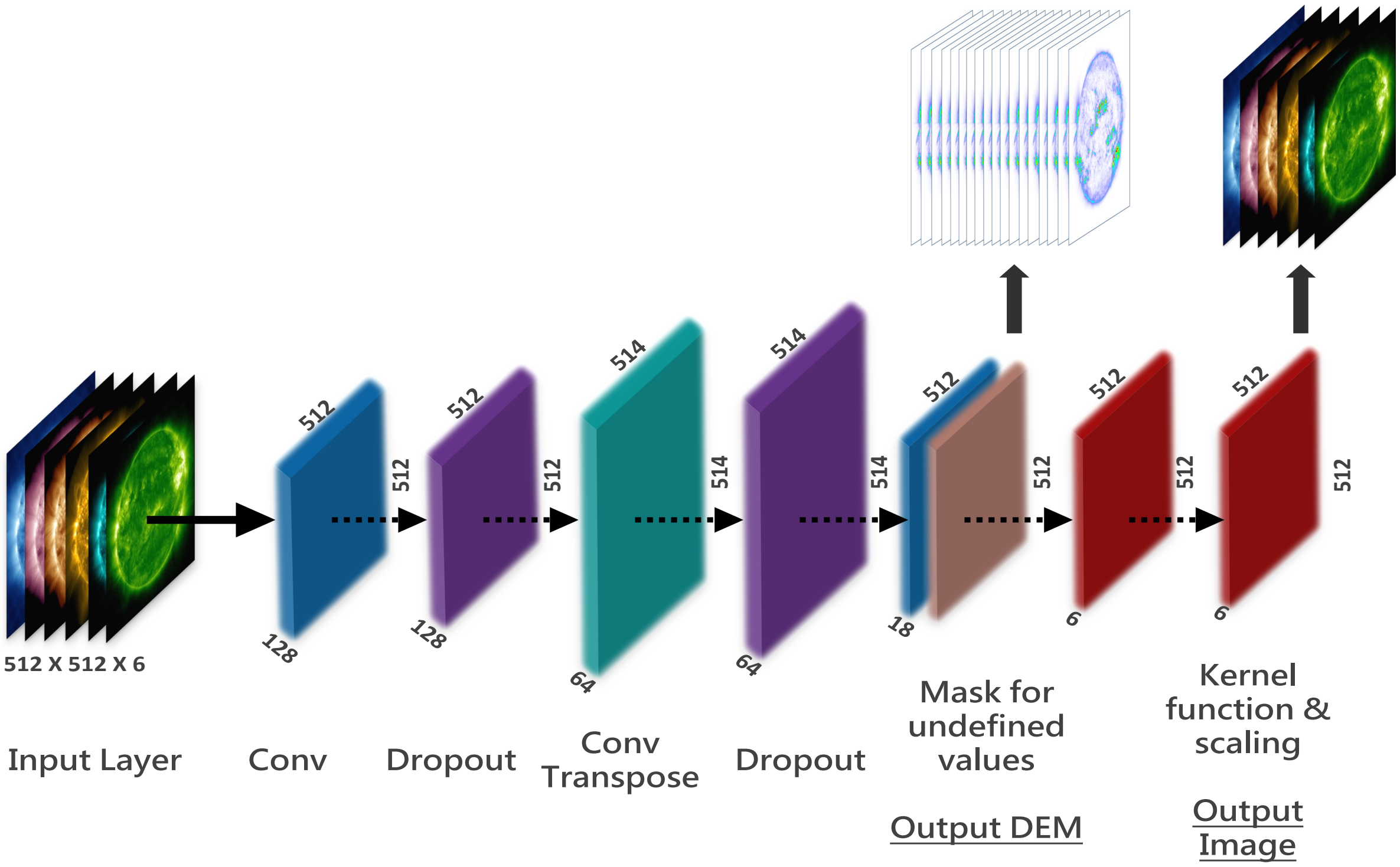

**Figure 2.** Bayesian model architecture with two target outputs.

This step ensures that the physical integrity of the original EUV data is preserved despite the application of learned regularization. A residual loss is calculated by comparing the reconstructed images with the original synthetic data used in training, thereby allowing the model to refine predictions iteratively.

The model architecture (Figure 2) consists of the following:

1. Two convolution layers with kernel sizes $1 \times 1$ and $3 \times 3$ to employ pixel-level and local spatial information.
2. Variational dropout layers after each convolution with a dropout rate of 0.3, to mitigate overfitting and estimate predictive uncertainty.
3. As the convolution layers decrease the output size, a transposed convolution layer with filter size $3 \times 3$ is added to upsample the feature maps, allowing the network to reconstruct DEM maps at the desired spatial resolution.
4. The L2 regularized method generates several undefined or nonphysical values both on and off disk, showing up as zero pixels. These are predominantly found in regions with negligible DEM contribution or when the response functions are weak. A mask layer filters out data points corresponding to undefined or missing DEM values in the targets, ensuring these are excluded from the loss computation during training. Removing outliers from training helps in numerical stability and optimization efficiency. The output from this layer forms the first network output.
5. A lambda layer allows us to specify an arbitrary mathematical function as a layer for a sequential model. A set of lambda layers computes the synthetic AIA observations by convolving the predicted DEM maps with the precomputed response function tensor, as defined in Equation (3). This constitutes the second output of the network and enables direct comparison with the training data.

### 3.3. Model Loss and Inference

We evaluate two losses for the model output DEM and image. For the DEM loss, we compute the mean absolute error on a per-pixel basis, producing a residual map and optimizing the summed loss across the temperature bins with AdamW (I. Loshchilov & F. Hutter 2017), providing $L_2$ constrained DEMs. AdamW is an optimization algorithm that extends Adam (adaptive moment estimation is a highly efficient, first-order gradient descent algorithm; D. P. Kingma & J. Ba 2017) by decoupling the weight decay regularization term from the parameters while computing the loss function, and instead applying the weight decay directly to the parameters after the gradient update.

An efficient DEM inversion process returns a stable solution, that is, the forward mapping onto the intensity avoids the propagation of errors. To avoid this propagation of errors, we compare the reconstructed image with the AIA observable. However, the observable $y_i$ given in Equation (3) is subject to photon noise and thus cannot be directly compared with the reconstructed image. To evaluate the image maps, we thus employed a noise-stabilized loss function that optimizes an output mean along with a noise estimate.

A. Kendall & Y. Gal (2017) propose a custom loss function to capture aleatoric uncertainty, or the inherent training data uncertainty, by fitting a Gaussian likelihood and tuning an observational noise parameter $\sigma$ as an input dependent variable. The loss function we use is given by

$$\mathcal{L}_{\rm NN}(\theta) = \frac{1}{N}\sum_{i=1}^{N} \frac{1}{2\sigma^2}\|\boldsymbol{y}_i - \boldsymbol{f}(\boldsymbol{x}_i)\|^2 + \frac{1}{2}\log\sigma^2, \quad (6)$$

where $y_i$ and $f(x_i)$ are our input and target outputs for $N$ data points. The first term contains the contribution from the target output and allows us to implicitly learn the variance $\sigma^2$ over the training data to predict the output image over multiple noise realizations, while the second term acts as a regularizer to prevent the model from predicting infinite variance.

For our problem, we utilize this loss function to capture the observable noise from the instrument as an aleatoric uncertainty measure. This, however, is calculated on the second network output and cannot be directly added to give a total uncertainty (as shown in Equation (A6)). This approach allows us to reconstruct images with noise and calls for more realistic results. We note that A. Kendall & Y. Gal (2017) define $\sigma^2(x)$ as the input dependent aleatoric uncertainty rather than a fixed noise parameter. For our approach, we have learned $\sigma^2$ as a standard noise prevalent over the images captured by the AIA, rather than an image-dependent predictable that requires a significantly more complex model architecture to capture.

The total loss function is defined as a weighted sum of these residual maps, with a higher weight (70%) assigned to the loss associated with the reconstructed image, and the remaining 30% attributed to the DEM output. This weighting prioritizes the fidelity of the reconstructed signal, ensuring that the network preserves the observed signal while learning to infer the underlying DEM distributions.

To obtain robust and uncertainty-aware predictions, we perform 100 stochastic forward passes through the network, with dropout enabled at inference time. Each pass constitutes a sample from an approximate posterior distribution over the model parameters induced by the dropout, thereby enabling a Bayesian output estimation. The mean of this ensemble of predictions represents a posterior model output averaged over the epistemic noise, and the standard deviation captures the epistemic or model uncertainty. This averaged prediction can therefore be considered a more reliable and interpretable representation of the network's inference compared to any single stochastic realization.

## 4. Validation Tests

We evaluate the performance of our network architecture on a diverse set of synthetic Gaussian temperature distributions and actual AIA observations. We compare our results and the uncertainty estimates with the estimates and uncertainties from the regularized inversion technique.

### 4.1. Simulated Data

To assess the model's performance, we generate synthetic observations from analytically defined DEMs with Gaussian temperature profiles. We adopt Gaussian profiles in $\log T$ space, commonly used in the inversion of AIA data (C. Guennou et al. 2012a, 2012b), and define them as

$$\xi(T_j) = N_0 \sum_k \frac{\gamma_k}{\sqrt{2\pi}\,\sigma_{T,k}} \exp\left[\frac{-(\log T_j - \log T_{0,k})^2}{2\sigma_{T,k}^2}\right], \quad (7)$$

where $T_{0,k}$ denotes the centroid temperature, $\sigma_{T,k}$ is the standard deviation, and $N_0 = \int \xi(T)\, d\log T$. Here, the summation $k$ is for the number of Gaussians used and $\gamma_k$ represents the relative strength of each Gaussian ($\sum_k \gamma_k = 1$).

To simulate the intensity in each AIA bandpass, we compute the forward model by multiplying the synthetic DEM with the instrument response function ($K_i$) for each channel. A conversion factor of $T\ln(10) d\log T$ is applied to account for the logarithmic temperature binning used in the DEM representation. To produce realistic synthetic images, we incorporate photon and shot noise (P. Boerner et al. 2012; D. Yuan & V. M. Nakariakov 2012) into the calculated counts to get multiple realizations of the observed intensity and simulate the stochastic nature of observational data. The noisy photon count in the $i$th filter is modeled as $y_i + \alpha_i e_i$, where $y_i$ is the noise-free photon count in the $i$th filter, $e$ is the gain, and shot noise is added with a random variable $\alpha \sim \mathcal{N}(0, 1)$. Using this approach, we generate $512 \times 512$ realizations of the observable in each AIA channel, resulting in a set of realistic synthetic EUV images that serve as inputs for evaluating the performance and robustness of the network under realistic observational conditions.

#### 4.1.1. Plasma Diagnostics

To evaluate the performance of the Bayesian inference method, we generate a suite of Gaussian DEM models characterized by varying central temperatures and widths. We used a temperature grid ranging from $\log T_0 = 5.5$–$7.2$ with a smaller temperature bin spacing of 0.05, and widths ranging $\sigma = 0.05-0.6$ and an amplitude of $10^{21}$, that can be comparable to some active region emissions. It is important to note that this temperature grid is defined to generate the ensemble of ground-truth gaussian profiles and is different from the temperature data fed to the network for DEM inversion. These variations allow us to systematically assess the model's ability to recover DEMs across a range of thermal structures commonly encountered in coronal plasmas. We obtain DEMs across $n = 35$ temperature bins and noisy synthetic counts in $n = 6$ AIA channels using the response functions, which are then fed to the Bayesian network to obtain the inverted DEM distributions.

To assess the fidelity of the DEMs reconstructed from the synthetic data, we follow M. C. M. Cheung et al. (2015) and use their three metrics denoted as the first, second, and third moments of the DEM distribution. These metrics are defined as

$$\mathrm{EM} = \sum_j^n \xi_j, \quad (8)$$

$$\log T_{\rm DEM} = \mathrm{EM}^{-1}\left[\sum_j^n \xi_j \log T_j\right], \quad (9)$$

$$W_{\rm DEM}^2 = \mathrm{EM}^{-1}\left[\sum_j^n \xi_j (\log T_j - \log T_{\rm DEM})^2\right]. \quad (10)$$

The first moment, or the EM, is simply the total emission measure across all temperatures. $\log T_{\rm DEM}$ and $W_{\rm DEM}^2$ denote the DEM-weighted log temperatures and the thermal width of the distribution in $\log T$ space.

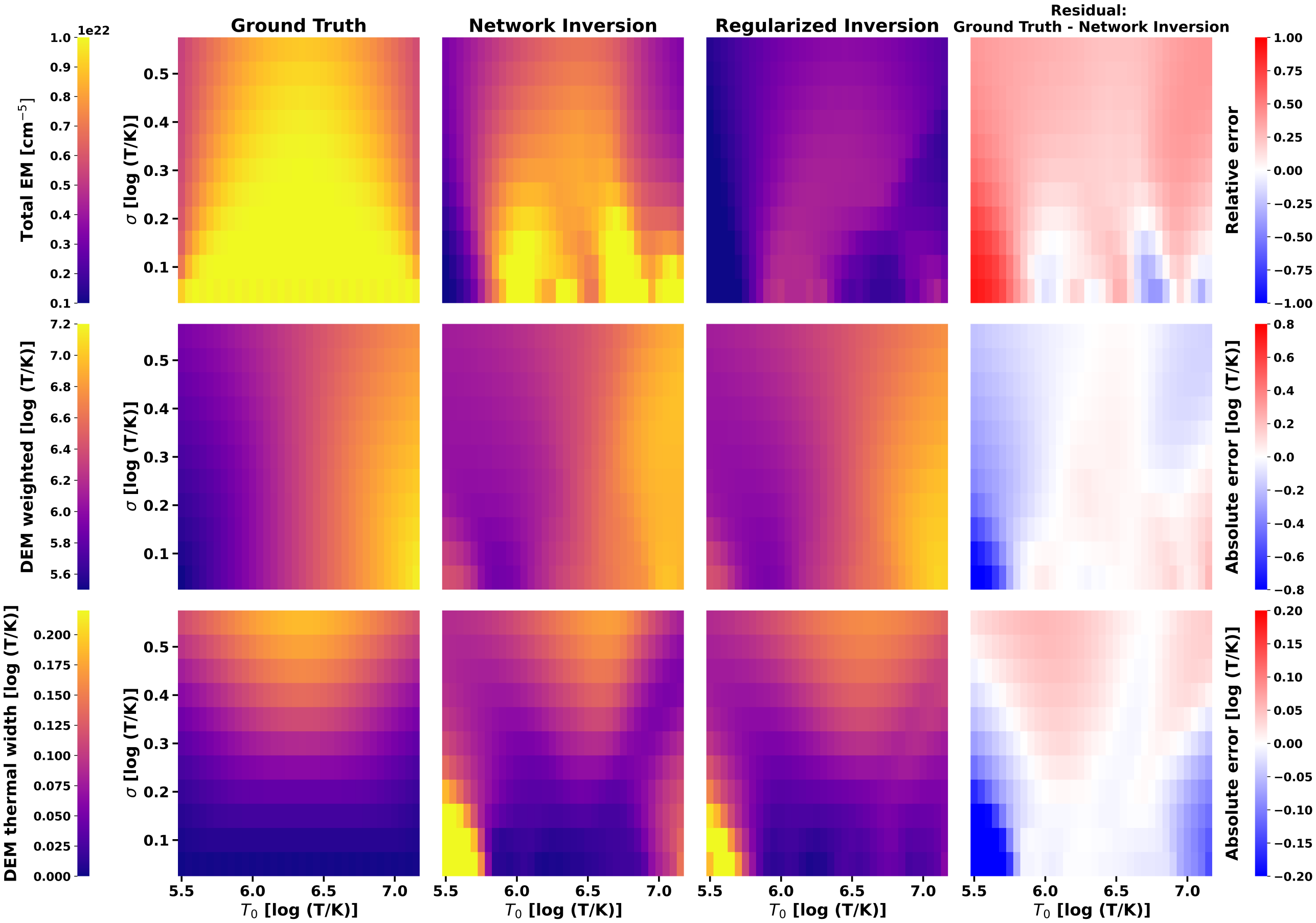

**Figure 3.** Results of validation tests for synthetic Gaussian DEM distributions in temperatures 1–10 million K. By adding Gaussian noise to the intensity obtained from these DEMs, we generate realistic AIA observables in six channels. The top, middle, and bottom rows represent the first, second, and third metrics for evaluating the fidelity of the inverted DEMs for the two methods.

Figure 3 compares these metrics for the ground truth with the inverted DEMs obtained by the network over the $T_0$, $\sigma$ space. The three rows correspond to the total emission measure, DEM-weighted mean temperature, and DEM thermal width, respectively. For each quantity, the left column shows the ground-truth values, the middle column the posterior mean of the inversion results, and the right column the residuals shown as the relative and absolute difference between ground-truth and inverted DEMs.

The inversion successfully reproduces the large-scale trends for the second and third metrics, but systematic biases are evident in the first metric and depend on both $T_0$ and $\sigma$. The total emission measure exhibits some discrepancies, with the inversion underestimating emission at temperatures $\log T < 5.8$, while the remaining temperature grid generally shows errors lying within 25%. By contrast, the DEM-weighted temperatures are robustly recovered over the parameter space with residuals confined within $\pm 0.8$ dex. The recovery of the DEM thermal width is intermediate in quality, where the error lies within $\pm 0.2$ dex, with poor performance at the lower temperature range. These results demonstrate that, while DEM inversions can accurately recover mean coronal temperatures and relative thermal structure, they are prone to errors due to the nature of the channel response functions. This emphasizes the need to practice caution while interpreting these DEMs.

In Figure 3, we observe various biases in the results from the DEM estimation procedure. However, we suggest that this underperformance can be explained by the lack of sensitivity from the combination of AIA passbands at certain temperatures. It is visible in Figure 1 that the combination of AIA channels used lacks sensitivity at low temperatures ($\log T < 5.8$) and also around $\log T \approx 6.5$. This leads to an underestimation of predicted plasma at the low temperatures. The worse performance at $\log T = 6.5$ is clear for narrow DEM profiles. This also explains the underestimation of material in the broad DEM profiles with significant plasma present at the temperatures with low sensitivity. This is supported by a comparison of our Figure 3 with Figure 2 in M. C. M. Cheung et al. (2015), which reveals the results from basis pursuit also have similar issues.

### 4.1.2. Single Gaussian DEMs

Using Dropout as a Bayesian approximation enables a quantification of the uncertainty within these inversions while ensuring time and computational efficiency. In Figure 4, we compare the results from the neural network to the direct estimates from the $L_2$ regularization method from I. G. Hannah & E. P. Kontar (2012) (i.e., the method used to generate the training data), through individual Gaussian DEM models, the recovered profiles, and the uncertainties obtained from the two methods. The results are displayed for four single, narrow

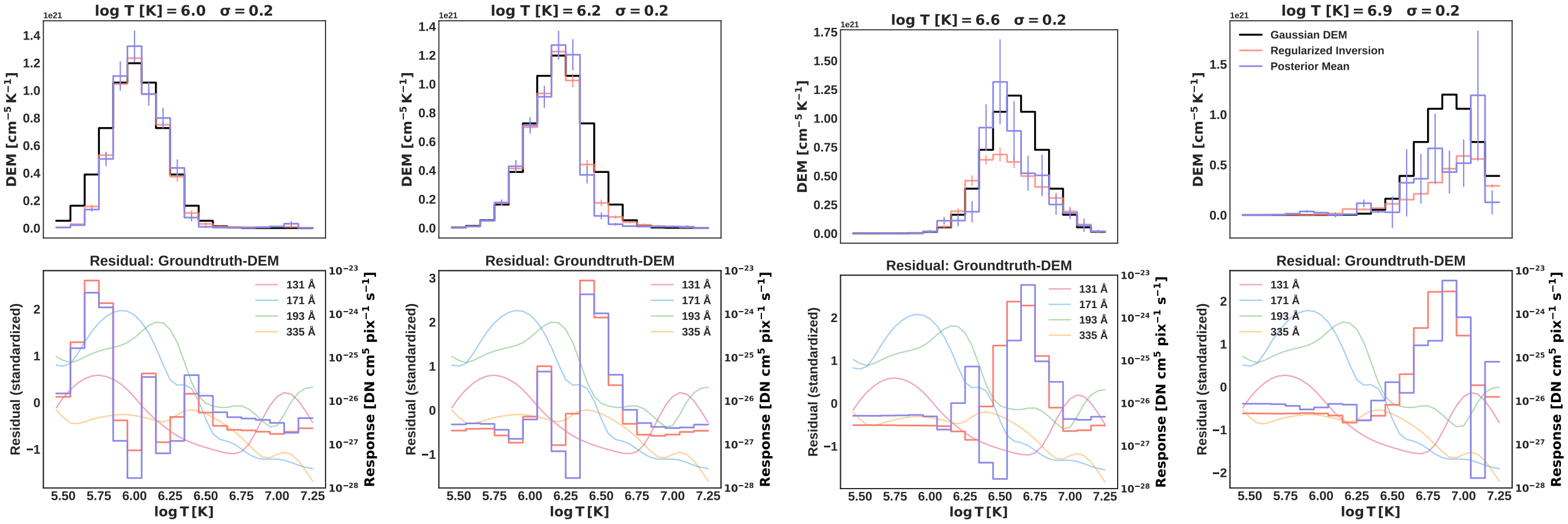

**Figure 4.** Recovery of synthetic single Gaussian DEMs with peak temperatures spanning $\log T_0 = 6.0$–6.9 and width $\sigma = 0.2$. The ground-truth profile is shown as a black curve. The Bayesian neural network posterior mean is plotted in purple, with the error bars representing the posterior standard deviation estimated via dropout sampling. The region between the error bars is a posterior credible interval. The $L_2$ regularized inversion result is shown in red along with the 68% confidence regions (note that these are fundamentally different in interpretation from the credible intervals). The bottom panel shows the standardized residuals from the ground truth along with the AIA response functions.

Gaussian DEM models in Figure 4 and two single, broad DEM model in Figure 5 with $N_0 = 6 \times 10^{20}$ and temperatures peaks varying from ($T_0 = 6.0-6.9$) with widths ($\sigma = 0.2, 0.4$).

In each panel, the ground-truth model (black curve) is compared with the posterior mean predicted by the dropout network (purple), where the error bars represent the associated standard deviation of the posterior (calculated from 100 stochastic forward passes). The reconstructions and corresponding errors obtained using the $L_2$ regularized inversion are shown in red. The $L_2$ regularized method generated several undefined or nonphysical values showing up as zero pixels, which are excluded from the analysis when computing the averaged emission measures. This ensures a consistent comparison between methods while preventing undefined values from biasing the reported emission statistics.

The posterior mean closely approximates the true DEM profile for narrow single Gaussian profiles for $\log T = 6.0$, 6.2 cases, accurately recovering sharp temperature peaks with standard deviation ranging from $10^{17}$ to $10^{19}$. Larger discrepancies are present between the prediction and the true DEM in regions where AIA has lower sensitivity, e.g., between $\log T = 6.5$ and 6.9. The standard deviation of the posterior is typically larger in these regions, and this reflects the uncertainty. The $L_2$ method shows similar results, but can underestimate the uncertainty at the temperatures with low AIA sensitivity. For the cases shown, it can be observed that the error bars associated with the $L_2$ regularized inversion are smaller and do not reflect the additional uncertainty in these regions well.[3] The differences between the two methods are indicated by the standardized residual plots, which also show four of the six response functions to indicate the temperature range with lack of sensitivity in AIA.

This performance can be contrasted with the lower fidelity of the network's estimates for broader DEM profiles lacking well-defined peaks (Figure 5). Although the network preserves the broad structure of the DEMs, we observe some underestimation of material with significant differences from the truth around $\log T \approx 6.5$. Notably, the dropout uncertainties appropriately reflect the limitations when the peak temperature of the true DEM is around $\log T \approx 6.5$, i.e., through increased uncertainty estimates (right panel in Figure 5). But when the true peak temperature is below $\log T \approx 6.5$, the estimated DEM misses the emission around $\log T \approx 6.5$ and does not represent the uncertainty well (left panel in Figure 5). In contrast, the regularized inversion method also predicts less emission around $\log T \approx 6.5$, but the estimated uncertainty is too small in both cases. Hence, this calls for some caution when dealing with inversion results showing broad profiles around the temperatures where AIA sensitivity is low.

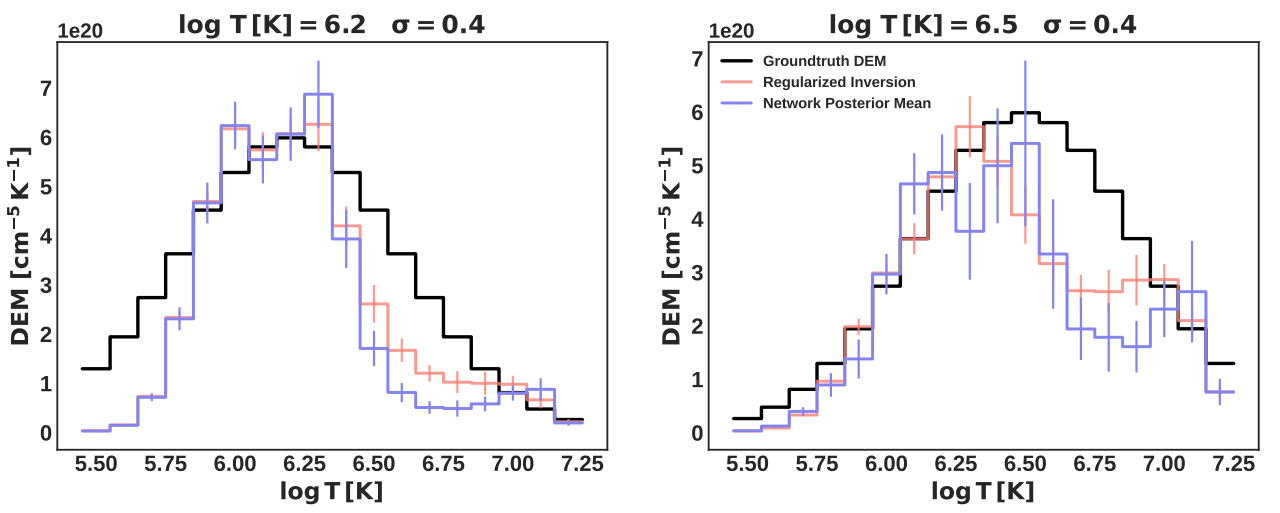

**Figure 5.** Same as Figure 4, but for broader profiles with $\sigma = 0.4$.

#### *4.1.3. Double Gaussian DEMs*

Figure 6 shows the reconstruction performance for representative double Gaussian DEM distributions, where two thermal components are combined with varying peak separations and widths by adding plasma emission with $N_0 = 6 \times 10^{20}$ and $N_0 = 10^{20}$ with $\sigma = 0.2$, 0.3 to single DEM profiles. Physically, these profiles may correspond to thermal conditions involving more complex structures that may arise from transient or spontaneous heating or cooling events in the solar atmosphere, leading to plasma emission across multiple temperature regimes along the LOS. These DEM profiles, while being more complex and sparsely represented in the training data, are recovered by the network with some underestimation in the thermal emissions around $\log T \approx 6.5$. Dropout resolves the components successfully, with higher error regions at the underestimated hot emissions. The peaks for temperatures below $10^6$ K exhibit a systematic shift toward higher temperature values. These limitations are consistent with the reduced sensitivity of the AIA instrument response functions at the extremes of the coronal temperature range.

[3] Note that the two intervals indicated by the error bars are fundamentally different in interpretation. A confidence interval indicates an interval calculated from repeated samples will contain the true, fixed parameter some percentage of the time (68% in the case shown in the figures). Conversely, a Bayesian credible interval identifies the narrowest range containing some percentage of the posterior probability. For the figures shown, this means there is a ∼68% probability the DEM value lies within the indicated range range

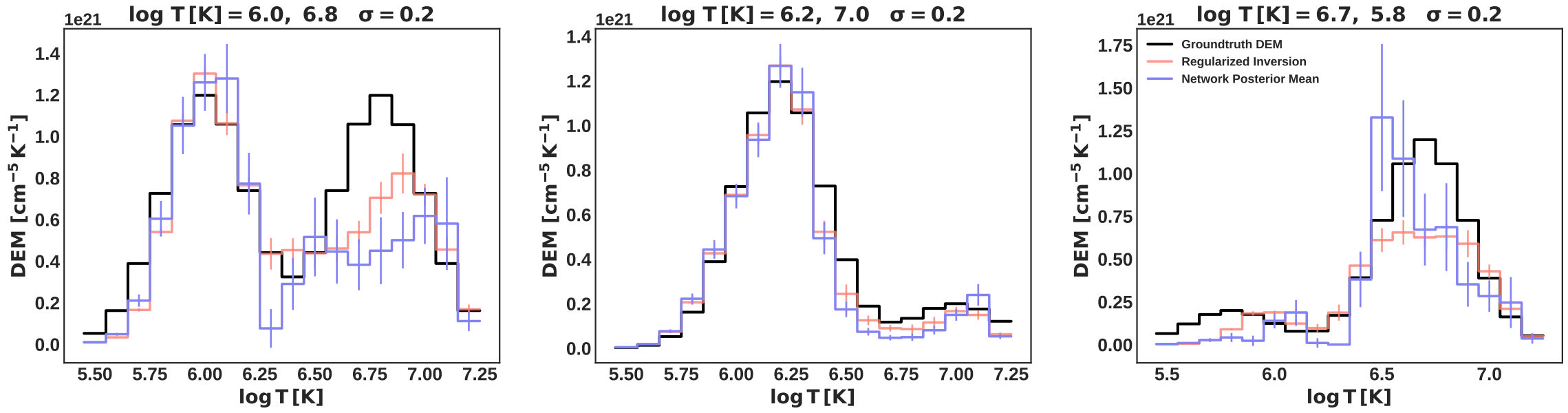

**Figure 6.** Reconstruction of synthetic double Gaussian DEM profiles with varying peak separations, amplitudes, and widths.

### 4.2. Application to AIA Data

The previous section shows us some valuable insights into the recovery of DEMs and the corresponding epistemic uncertainty through dropout regularization and how it supersedes other regularization methods in terms of performance and computational efficiency. We now demonstrate the performance of the network when estimating the DEMs from real AIA data.

Figure 7 presents the inverted DEM maps of the active region NOAA AR 13664 (R. Jarolim et al. 2024), a highly flare-productive region observed during the solar maximum phase in 2024 May (H. Hayakawa et al. 2025). The maps capture the thermal structure of the region taken at 2024-05-07T00:00. The displayed DEMs correspond to the posterior mean obtained from 100 stochastic forward passes of the observed AIA intensity through the dropout network. In principle, as each forward pass is considered a draw from the posterior, one can also obtain a valid posterior credible interval for the resulting sum over the bins (further discussion in Section 5.3).

Figure 7 spans the temperature range $\log T \in [5.75, 7.25]$, enabling a comprehensive view of the multithermal plasma, revealing significant emission in the mid to high temperatures, indicating the presence of hot plasma components associated with active region heating and flaring processes.

As mentioned above, the proposed network can deliver samples from an approximate posterior. Computing the posterior mean over these samples is therefore equivalent to marginalizing over the predictive noise resulting in smooth output maps. This property is useful for denoising images. Figure 8 compares the AIA instrument observations of the NOAA AR 13664, along with a quiet Sun region from the same time stamp at 193 Å, to a single prediction of the output image by the dropout network and an average of 100 predictions. Each set of predicted images (i.e., one for each AIA channel) arises from passing the posterior sample DEM through the response function. The posterior images largely preserve the structure and magnitude of the original data, demonstrating that the DEM inversion is consistent with the AIA observables but can lead to some loss of signal in the hot flaring regions. Averaging 100 stochastic predictions sampled from the approximate posterior also yields a noticeably smoother and more stable reconstruction compared to any single realization, and thus can be regarded as a reliable and physically consistent representation of the underlying emission measure distribution. To visualize differences, we compute the difference of the AIA image and the posterior mean, and standardize it by the per-pixel uncertainty calculated from expected data noise (right panels in Figure 8). Although not shown, we confirmed the reasonably faithful reconstruction of images by comparing the circularly averaged 2D power spectrum for the original image and posterior mean image. The two show differences in the higher frequencies that suggest noise suppression. We observe a consistent performance for the quiet Sun region across the other channels. However, there is some displacement of the signal in the flaring active region between channels. This could be due to the predicted DEM moving the differential emission between the temperature bins and resulting in signal redistribution.

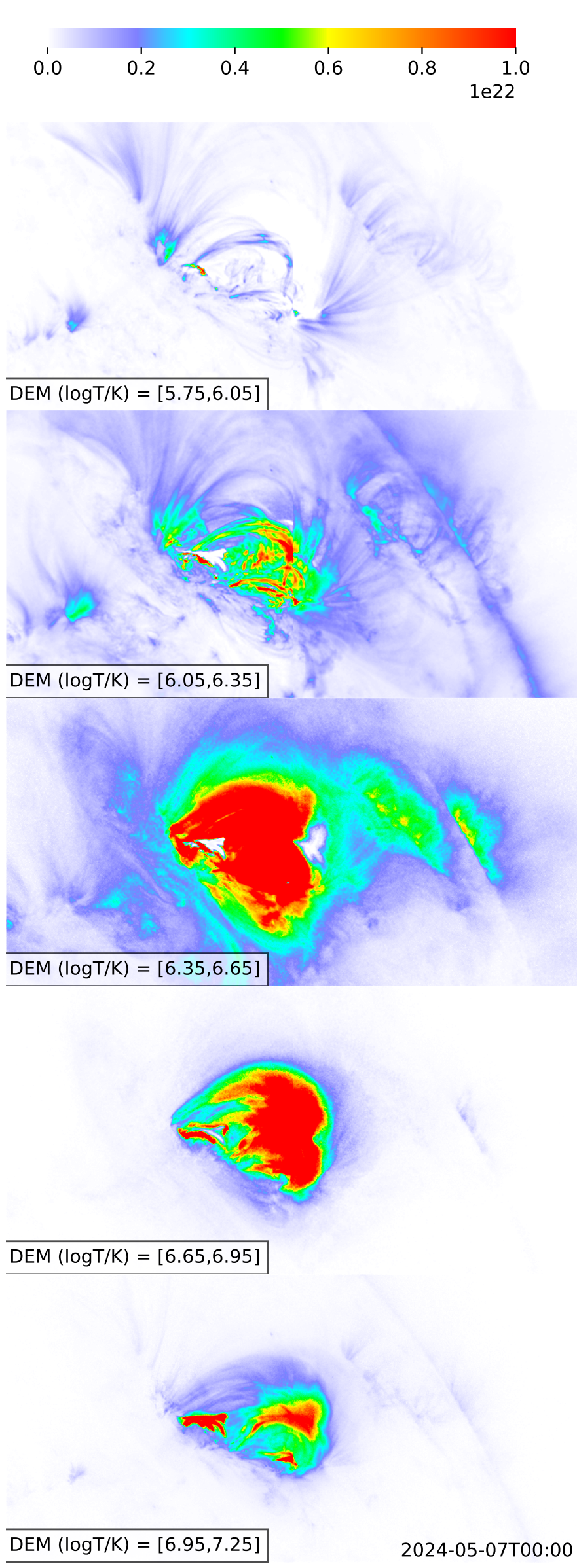

**Figure 7.** Posterior mean of DEM estimates from the dropout network for the active region NOAA AR 13664 over a flaring activity captured at 2024-05-07T00: 00. The emission measures are captured for $\log T \in [5.75, 7.25]$ within temperature intervals of 0.1 K and then accumulated to provide the total emissions in the intervals given in the bottom left of each plot.

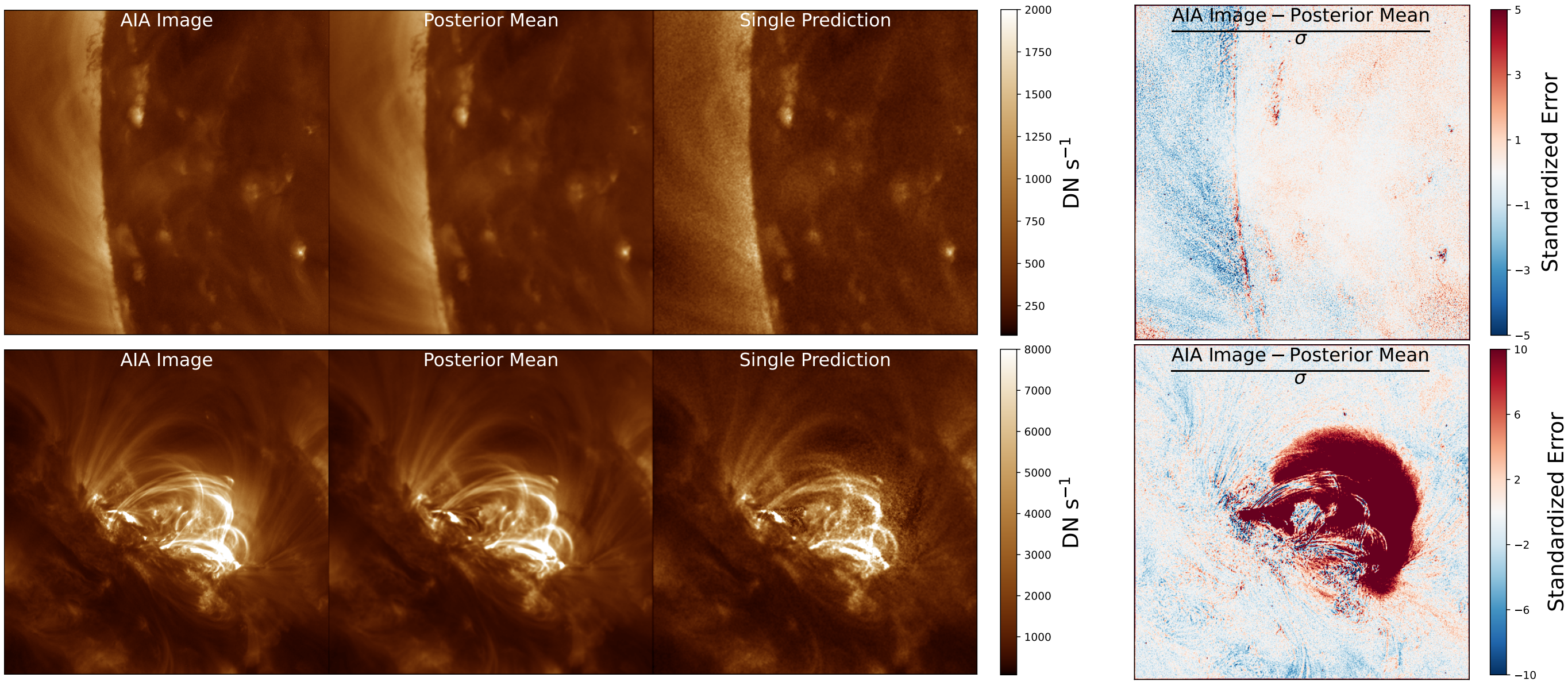

**Figure 8.** Comparing a quiet Sun (top) and active region NOAA AR 13664 (bottom) captured by the AIA (left panel) with the reconstructed image by the network in the 193 Å channel. The right panel shows a single prediction or a sample from the approximate posterior, and the middle panel shows the posterior mean estimated from 100 predictions. A final panel (hard right) shows the subtracted residuals between the AIA image and the posterior mean, standardized by the per-pixel estimated uncertainty.

Careful inspection of the posterior mean image also reveals that the current architecture used for the network actually results in a blurring of the features. This is somewhat typical of denoising processes (although with some exceptions, e.g., C. E. DeForest 2017). The potential denoising aspect is not a principle focus here, and it is currently unclear what type of network architecture would/could alleviate the blurring and signal redistribution. However, improving the denoising aspect is one avenue that further methodological improvements could focus on.

## 5. Discussion

### 5.1. Nature of the Uncertainty Estimates

A key motivation for the present work was to provide per-pixel uncertainty estimates for the recovered DEM. It is important to discuss how these relate to the uncertainty estimates provided by the regularized inversion of I. G. Hannah & E. P. Kontar (2012). Their uncertainty estimates are derived from a frequentist perturbation approach, based upon assuming a Gaussian distribution for the data noise. The input data are perturbed within the proposed noise bounds, and estimates are given for how much the solution varies through an MC approach. This is a purely aleatoric approach and can be viewed as a sensitivity analysis of the model to data noise. I. G. Hannah & E. P. Kontar (2012) acknowledge that the exact statistics of the observational errors are unknown, noting that the resulting DEM uncertainties may themselves be non-Gaussian.

The uncertainty estimates produced by the neural network are qualitatively different. By retaining dropout at inference time and performing stochastic forward passes, we obtain an ensemble of DEM predictions whose empirical distribution approximates the epistemic component of the predictive uncertainty, that is, the uncertainty in the learned mapping arising from uncertainty in the network weights. This is not sensitivity to input data noise, but rather a characterization of where the network is uncertain about how to perform the inversion, reflecting the limits of what can be resolved from the training data. For a given pixel, a broad empirical distribution across forward passes indicates that the network has not converged to a consistent solution in that region of input space, whether due to limited training data or genuine ambiguity in the mapping.

It follows that these two uncertainty quantities are not interchangeable and their relationship is not straightforward. A pixel where the regularized inversion reports large vertical error bars (indicating high sensitivity to data noise) might not coincide with a pixel where our network reports large epistemic uncertainty, and vice versa. The two uncertainty sources are, in principle, complementary. One reflects the noise properties of the instrument, while the other reflects the completeness and consistency of the learned inversion. A full uncertainty budget for the DEM would ideally combine both; the present work characterizes only the epistemic component.

### 5.2. Shared Properties

An important point of discussion is that the present network was trained on DEM solutions produced by the regularized inversion of I. G. Hannah & E. P. Kontar (2012). The network has therefore learned to approximate that method's mapping from AIA channel intensities to DEM solutions. The training labels carry the assumptions and systematic characteristics of the regularized inversion. In particular, any systematic bias in the I. G. Hannah & E. P. Kontar (2012) solutions (as noted by, e.g., P. Massa et al. 2023) will be inherited by the network as a bias in the training data rather than corrected by the learning process. The same can said of the DeepEM method and its relationship to the basis pursuit solutions used as training data.

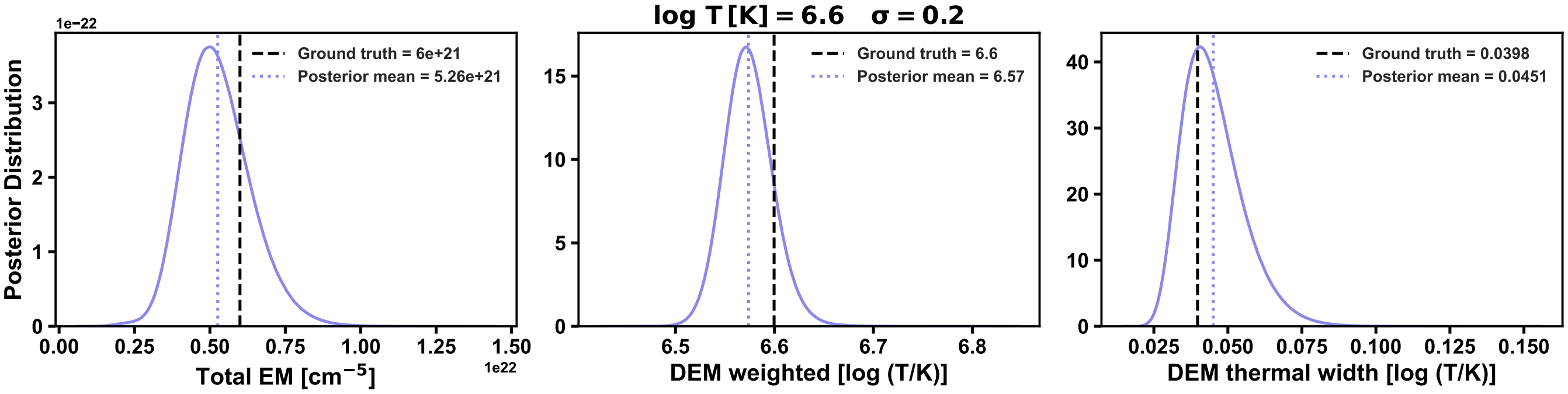

**Figure 9.** Posterior distributions for the total emission measure, DEM-weighted temperature, and thermal width for a DEM profile with peak $\log T = 6.6$ and $\sigma = 0.2$. The plasma diagnostics are computed individually for each of the 100 stochastic passes from the network and plotted as a distribution. The dotted black line denotes the ground truth for the DEM profile, and the dotted line is the posterior mean.

This has implications for the interpretation of the epistemic uncertainty estimates. The stochastic forward passes characterize the network's uncertainty about how to reproduce the I. G. Hannah & E. P. Kontar (2012) inversion, not uncertainty about the true DEM. A region where the network is highly confident need not be a region where the underlying inversion is reliable. It may simply be a region where the mapping from AIA intensities to the regularized inversion solutions is well constrained and consistent in the training data. On the other hand, high epistemic uncertainty at a given pixel can be interpreted as indicating that the training data did not provide sufficient coverage to determine a stable approximation to the reference inversion in that region of observation space.

### 5.3. Posterior Distributions for Derived Quantities

The significance of the MC process employed in the network predictions extends beyond just computing the model uncertainties. A further advantage of the present approach is the capacity to perform principled uncertainty propagation to derived plasma diagnostics. Because the stochastic forward pass ensemble constitutes samples from the approximate predictive posterior over the DEM, any deterministic transformation of the DEM, such as the emission-measure-weighted temperature, column emission measure, or thermal energy density, can be computed independently for each forward pass. The sample values can be then used to construct an empirical distribution that approximates the so-called push forward of that posterior under the transformation. This enables posterior probability statements about derived quantities, e.g., credible intervals. An example of the posterior distributions computed for the plasma diagnostics is shown in Figure 9, calculated for a DEM profile peaking at $\log T = 6.6$ and $\sigma = 0.2$ and using 100 forward passes from the network.

This is not straightforward to do in a frequentist modeling approach (e.g., basis pursuit or $L_2$ regularization), and highlights the benefit of utilizing a Bayesian framework. The uncertainties given by the method outlined in I. G. Hannah & E. P. Kontar (2012) do not lead to a posterior in any principled sense. Hence, the present framework extends uncertainty quantification beyond the DEM itself to derived coronal plasma diagnostics. This, of course, comes with the caveat that all such posterior distributions are conditioned on the network having learned to approximate the $L_2$ regularized inversion, and they reflect epistemic uncertainty in that learned mapping rather than uncertainty with respect to the true plasma state.

### 5.4. Other Benefits of the Current Approach

Despite the caveats above, the present method offers a practical and significant advantage over the regularized inversion for large-scale solar data analysis. The regularized inversion of I. G. Hannah & E. P. Kontar (2012) requires an independent matrix inversion for each pixel, and while computationally fast relative to MCMC-based methods (e.g., V. Kashyap & J. J. Drake 1998), it scales linearly with the number of pixels. While prior deep learning efforts successfully reduced these computational burdens (P. J. Wright et al. 2019), they followed a frequentist approach that could not provide an estimate of uncertainty on the predictions.

The present network, once trained, performs the inversion as a single forward pass through a convolutional architecture that processes a 512 × 512 cutout from a full-resolution AIA image simultaneously in a matter of a couple seconds. For uncertainty-aware predictions, our method yields a speedup of approximately 4 minutes for 100 forward passes relative to the regularized inversion method on an AMD Ryzen 3 PRO 3200g processor. The network can further immensely speed up the process of obtaining DEMs for full-resolution images in about 70 minutes by computing individual cutouts and stacking them, significantly reducing the computational speed compared to the $L_2$ method of 150 minutes run in parallel, along with temperature resolution on a smaller scale. This enables uncertainty-quantified DEM reconstruction at full AIA cadence and resolution, which is not feasible with traditional inversion approaches.

Furthermore, the epistemic uncertainty maps produced are useful in themselves. Regions of high epistemic uncertainty are likely to correspond to observational conditions that were underrepresented in the training data, the learned mapping is poor, and/or such temperature regions are not well constrained. This latter effect is seen in the synthetic examples where AIA shows limited sensitivity at temperatures around $\log T \sim 6.5$ (Figures 4 and 5). These maps therefore serve as a practical diagnostic for the reliability of the reconstruction on a per-pixel basis, allowing users to identify regions where the network solution should be treated with caution.

Finally, the standard mathematical approaches, such as those using regularization, can often produce undefined or physically illegitimate negative emission values. The network addresses this issue and is able to provide physically meaningful DEMs at all pixels. This effect was also observed in the DeepEM solutions.

## 6. Conclusion

In this study, we presented a Bayesian deep learning framework for estimating plasma emissions from the solar corona using observations from the SDO/AIA instrument in optically thin EUV wavelengths. By incorporating variational dropout into the network, it serves as a practical approximation to Bayesian inference. The proposed model generates stable and physically plausible DEM reconstructions. Furthermore, repeated stochastic forward passes with dropout retained at inference time yield an ensemble of predictions whose empirical distribution approximates the epistemic component of the predictive posterior, reflecting model uncertainty due to uncertainty in the network weights. The ensemble of predictions can then be used to estimated uncertainty on derived quantities. More formally, the empirical distribution of, say, emission measure, obtained by applying Equation (1) to each stochastic forward pass, approximates the propagated epistemic uncertainty in that quantity, inherited from model weight uncertainty via MC dropout. Estimating such uncertainty is an essential step toward constraining the reliability of coronal temperature diagnostics derived from machine learning models. However, it is worth noting that other sources of uncertainty also remain, e.g., due to the atomic calculations used for the response functions, which are not accounted for in this process.

It is also worth emphasizing again that, in our approach, the true DEMs are not in the training data, only one particular regularized approximation to it is. The network has therefore learned to approximate a solution to an already approximated problem (the same caveat also applies to DeepEM; P. J. Wright et al. 2019). Future improvements on this aspect could be found by training a network on data from MHD simulations, from which we can obtain the true DEM distributions and create synthetic, noisy data that represent SDO/AIA images. The model and weights are given in the repository Bayesian Deep Learning approach to DEM estimation.[4]

## Acknowledgments

All authors acknowledge the UK Research and Innovation (UKRI) Science and Technology Facilities Council (STFC) for support from grant No. ST/W006790/1. This study has been supported by the STFC Centre for Doctoral Training in Data Intensive Science (NUdata), as a collaboration between Northumbria and Newcastle Universities. R.J.M. is supported by the UKRI Future Leaders Fellowship (RiPSAW MR/T019891/1 and MR/Z000289/1).

*Facility:* SDO.

[4] https://github.com/nikitabalodhi1/bayesian_dem

*Software*: Numpy (C. R. Harris et al. 2020), SciPy (P. Virtanen et al. 2020), Matplotlib (J. D. Hunter 2007), Tensorflow (M. Abadi et al. 2015).

## Appendix
## Variational Inference and MC Dropout

The exact posterior predictive distribution (Equation (4)) requires marginalizing over the true posterior $p(\boldsymbol{w}|D)$, which is analytically intractable for neural networks of practical depth and width. Following Y. Gal (2016), dropout training admits a variational inference interpretation by introducing an approximate posterior $q(\boldsymbol{w})$ over the network weights. The predictive distribution under this approximation becomes

$$p(\boldsymbol{y}^*|\boldsymbol{x}^*, D) \approx \int p(\boldsymbol{y}^*|\boldsymbol{x}^*, \boldsymbol{w})q(\boldsymbol{w})\, d\boldsymbol{w}. \tag{A1}$$

The variational distribution $q(\boldsymbol{w})$ is a mixture of two delta functions for each network weight, placing probability mass either at zero (the unit is dropped) or at the trained weight value $\hat{w}_i$ (the unit is retained):

$$q(w_i) = p_i\, \delta(w_i) + (1 - p_i)\, \delta(w_i - \hat{w}_i), \tag{A2}$$

where $p_i$ is the dropout probability for unit $i$. Each stochastic forward pass therefore draws a sample $\boldsymbol{w}_t \sim q(\boldsymbol{w})$, corresponding to a randomly selected binary mask applied to the weight matrices.

Variational inference requires minimizing the Kullback–Leibler (KL) divergence between the approximate and true posteriors,

$$\mathrm{KL}[q(\boldsymbol{w}) \mid p(\boldsymbol{w}|D)], \tag{A3}$$

which is equivalent to maximizing the evidence lower bound on the log marginal likelihood. For this to correspond to standard dropout training, a prior distribution on the network weights must also be specified. Y. Gal & Z. Ghahramani (2016) show that a Gaussian prior on the weights satisfies this condition and directly implies $L_2$ regularization (weight decay) in the training objective. Under this prior, the $L_2$ regularization coefficient $\lambda$ is related to the model hyperparameters by

$$\lambda = \frac{\ell^2(1 - p)}{2N\tau\sigma^2}, \tag{A4}$$

where $p$ is the dropout probability, $N$ is the number of training samples, $\tau$ is the model precision (inverse of the observation variance), $\sigma^2$ is the prior variance on the weights, and $\ell$ is a length-scale parameter arising in the Gaussian process interpretation of the infinite-width network limit (Y. Gal 2016).

Once the neural network with dropout has been trained, the predictive distribution of the model on test data can then be given by an MC estimate. While making predictions, the model can predict multiple distinct values of $\boldsymbol{y}^*$ for each value of $\boldsymbol{x}^*$, due to randomized MC dropout of the network parameters, and subsequently, a distinct conditional probability distribution, $p(\boldsymbol{y}^*|\boldsymbol{x}^*, \boldsymbol{w}_t)$, for each input. Hence the MC approximation for the predictive distribution is

$$p(\boldsymbol{y}^*|\boldsymbol{x}^*, D) \quad \approx \frac{1}{T}\sum_{t=1}^{T} p(\boldsymbol{y}^*|\boldsymbol{x}^*, \boldsymbol{w}_t), \tag{A5}$$

where $T$ is the number of forward passes through the network, and $\boldsymbol{w}_t \sim q(\boldsymbol{w})$. The posterior predictive mean is then given by

$$\mathbb{E}_q(\boldsymbol{y}^*) \approx \frac{1}{T}\sum_{t=1}^{T}\hat{\boldsymbol{y}}(\boldsymbol{x}^*, \boldsymbol{w}_t) = \hat{\mu}_{\boldsymbol{y}}.$$

Given the posterior predictive mean $\hat{\boldsymbol{\mu}}_y$ from Equation (A5), the total predictive variance over $T$ stochastic forward passes decomposes as

$$\mathrm{Var}_q(\boldsymbol{y}^*) \approx \left(\frac{1}{T}\sum_{t=1}^{T}\hat{\boldsymbol{y}}(\boldsymbol{x}^*, \boldsymbol{w}_t)^T\hat{\boldsymbol{y}}(\boldsymbol{x}^*, \boldsymbol{w}_t) - \hat{\boldsymbol{\mu}}_y^2\right) + \tau^{-1}. \quad (A6)$$

The first term is the variance of the MC ensemble of predictions, reflecting uncertainty in the network weights $\boldsymbol{w}_t \sim q(\boldsymbol{w})$. This is the *epistemic* component, which is reducible, in principle, with additional training data. The second term $\tau^{-1}$ is the *aleatoric* component, representing irreducible noise intrinsic to the observations themselves.

## ORCID iDs

N. Balodhi https://orcid.org/0009-0008-3777-7557
R. J. Morton https://orcid.org/0000-0001-5678-9002